%% file: main_camera_ready.tex
\documentclass[conference,compsoc]{IEEEtran}
\IEEEoverridecommandlockouts
\usepackage[T1]{fontenc}
\usepackage{graphicx}
\usepackage{listings}
\usepackage{tabularx}
\usepackage{multirow}
\usepackage{array}
\usepackage{booktabs}
\usepackage{xurl}
\usepackage{cite}
\usepackage[table]{xcolor}
\usepackage{amsmath}
\usepackage{amssymb}
\usepackage{ragged2e}
\usepackage[most]{tcolorbox}
\usepackage[hidelinks]{hyperref}

\newcolumntype{Y}{>{\RaggedRight\arraybackslash}X}
\begin{document}
\title{The Devil Wears Markdown: Detecting Malicious Intent Hidden in AI Agent Skill Instructions through Hybrid Static, Dynamic, and Agentic Analysis}
\author{
\IEEEauthorblockN{
Thanh M. Truong-Le\IEEEauthorrefmark{1},
Amber Bennoui\IEEEauthorrefmark{2},
Hoang-Ly Nguyen\IEEEauthorrefmark{3},
Gia-Thang Ho\IEEEauthorrefmark{4},\\
Thai-Minh Truong\IEEEauthorrefmark{4},
Duc-Ly Vu\IEEEauthorrefmark{5}\textsuperscript{1}%
\thanks{\textsuperscript{1}\,Corresponding author:
Duc-Ly Vu (e-mail: lyvd@tdmu.edu.vn).}
}
\IEEEauthorblockA{
\IEEEauthorrefmark{1}University of Information Technology, Vietnam\\
thanhtlm.19@grad.uit.edu.vn\\
\IEEEauthorrefmark{2}Jiffy Labs, USA\\
amber@jiffylabs.ai\\
\IEEEauthorrefmark{3}Eastern International University, Vietnam\\
ly.nguyen@eiu.edu.vn\\
\IEEEauthorrefmark{4}Ho Chi Minh City University of Technology (HCMUT),
Vietnam National University, Vietnam\\
thang.ho261104@hcmut.edu.vn; thaiminh@hcmut.edu.vn\\
\IEEEauthorrefmark{5}Thu Dau Mot University, Vietnam\\
lyvd@tdmu.edu.vn
}
}
\maketitle

\begin{abstract}
\input{abstract.tex}
\end{abstract}


\input{introduction.tex}

\input{background_related_work.tex}
\input{methodology.tex}

\input{experiment_setup.tex}
\input{result.tex}
\input{Threats_to_Validity.tex}
\input{conclusion.tex}

\bibliographystyle{IEEEtran}
\bibliography{ref_shortened}

\appendices
\input{appendix.tex}
\end{document}

%% file: abstract.tex
Agentic AI ecosystems increasingly rely on reusable agent skills
that combine executable code, natural-language instructions,
configuration artifacts, and external tool integrations.
These hybrid artifacts create software supply-chain risks
because malicious intent may be distributed across instructions,
code, metadata, and runtime behavior, making isolated alerts
difficult to interpret.

We evaluate three open-source scanners on a held-out test set
of 140 skills comprising 80 malicious and 60 benign samples,
including Cisco Skill Scanner with and without its
optional LLM analyzer. The default Cisco configuration achieves
a recall of 0.91 with 33 false positives among the 60 benign
skills, while the LLM-enabled configuration using
Qwen3:8B achieves a recall of 0.85 with 40 false positives.

We then present a framework that combines static findings,
safety-constrained sandbox evidence, and LLM-based reasoning
to assess malicious intent. The framework tracks the origins
of findings and relates evidence across analysis layers.
In a within-system evidence-source ablation using the same
reasoning backend, adding dynamic evidence to skill-content
reasoning increases F1 from 0.70 to 0.85. Combining static
and dynamic evidence further increases F1 to 0.89, with
five false positives among 60 benign skills. Dynamic evidence
provides the largest observed improvement.

The Full Pipeline detects 68 of 80 malicious skills,
achieving 0.88 accuracy, 0.93 precision, and 0.85 recall.
It matches the observed recall of Cisco Skill Scanner with
LLM while reducing false positives from 40 to five.
This system-level comparison does not isolate model
or architecture effects.

%% file: introduction.tex
\section{Introduction}

\begin{figure*}[!t]
\centering
\includegraphics[width=0.95\textwidth]{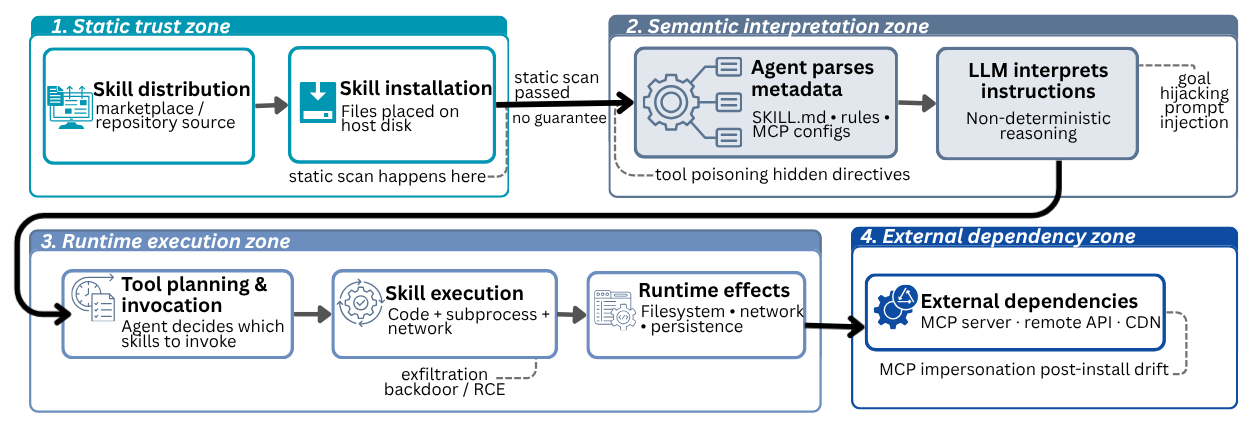}
\caption{Install-to-execution trust chain in agent skill ecosystems.}
\label{fig:install-to-execution}
\end{figure*}

Large language model (LLM)-based agents increasingly rely on external extensions to perform specialized tasks. In ecosystems such as Claude Code, these extensions are packaged as \emph{skills}, which may contain natural-language instructions, scripts, configuration files, and supporting resources that are loaded when relevant~\cite{anthropicSkills,anthropicSkillsGithub}. Although skills support software engineering, deployment, API integration, and workflow automation, they also introduce security risks because both instructions and executable artifacts can influence future agent behavior. Figure~\ref{fig:install-to-execution} illustrates the install-to-execution lifecycle of an agent skill, from developer publication to runtime effects, and highlights how trust assumptions change across stages. Different points in this lifecycle create multiple trust surfaces that cannot be fully validated at any single stage~\cite{liuMaliciousSkillsWild2026}. Unlike traditional software packages, agent skills are interpreted and executed through an AI agent, which means that malicious behavior may arise not only from code execution but also from hidden instructions, excessive agency, tool poisoning, permission abuse, or mismatches between the documented purpose and the actual behavior~\cite{krishna2026clawdpwned,liuMaliciousSkillsWild2026,virustotal2026openclaw}.

Recent studies have shown that agent skills represent an emerging attack surface. Liu et al.~\cite{liuAgentSkillsWild2026} reported vulnerabilities such as prompt injection, data exfiltration, privilege escalation, and supply-chain risks in large-scale public skill ecosystems. A follow-up study further showed that confirmed malicious skills often contain multiple vulnerabilities spanning several stages of an attack chain, including data theft and agent hijacking~\cite{liuMaliciousSkillsWild2026}. These findings suggest that malicious behavior is rarely captured by a single suspicious keyword or command. Instead, attacks are often distributed across instructions, scripts, configurations, runtime actions, and external communications.

Liu et al.~\cite{liu2026trust} show that malicious agent skills do not always contain explicit malware indicators. They propose a technique called \textit{Dependency Steering}, which manipulates coding agents into recommending attacker-selected packages during otherwise benign programming tasks. Lin et al.~\cite{lin2026phantomskill} demonstrate a technique for concealing malicious behavior in auxiliary skill resources by rewriting explicit payloads as vulnerability-shaped code. These attacks show that skill-induced harm may manifest through downstream agent decisions, trigger-dependent behavior, or code that resembles an ordinary software vulnerability.

Industry and academia have proposed scanners for malicious agent skills. Cisco Skill
Scanner~\cite{cisco_skill_scanner}, for example, combines signatures, code and bytecode
inspection, static behavioral analysis, and optional LLM analysis. However, existing
tools lack systematic evaluation on real-world skills, particularly regarding detection
performance and false-positive noise. We therefore ask the following research question:
\begin{itemize}
    \item \textbf{RQ1:} How do existing open-source scanners perform on a dataset of malicious and benign agent skills?
\end{itemize}

To answer this research question, we evaluate existing malicious-skill scanners on a set of malicious and benign agent skills. 
For scanner configurations evaluated using alert counts,
we classify a skill as malicious when its number of reported
alerts reaches or exceeds a selected threshold. These thresholds
are selected exclusively on the development split and applied
unchanged to the held-out test set. Configurations that provide
a native binary verdict are evaluated using that verdict directly.
Here, the development split supports detector configuration
and threshold selection rather than LLM weight training.
We compare the tools in terms of detection performance, ranking quality, and false-positive noise. This evaluation provides a baseline that quantifies the limitations of existing scanners and motivates the design requirements of our framework.


Agent skills differ from traditional open-source packages in that they contain code, such as Python scripts, as well as natural-language text, such as instructions in \texttt{SKILL.md}. Static analysis techniques examine an agent skill without
executing it. Rule-based techniques can identify suspicious
code patterns, but interpreting these findings requires
context about the skill's intended behavior. Dynamic analysis complements static analysis by providing runtime evidence of actions such as
credential access, outbound data transfer, and persistence
attempts that may be difficult to confirm from skill artifacts
alone~\cite{cisco_skill_scanner, liuMaliciousSkillsWild2026}. Interpreting static findings alongside runtime evidence
can help assess whether suspicious operations are
consistent with a skill's stated purpose. This motivates
our second research question:


\begin{itemize}
\item \textbf{RQ2:} How can static analysis, dynamic analysis, and LLM-based reasoning be combined to correlate evidence and improve the detection of malicious agent skills?
\end{itemize}

To answer this research question, we propose a framework that combines static analysis, dynamic analysis, and agentic reasoning to detect malicious agent skills. We evaluate the framework on a curated dataset of malicious and benign skills and compare its performance with that of existing scanners.

In summary, this paper makes three contributions:

\begin{itemize}
    \item We benchmark three open-source scanners on a common
    development/test split, including Cisco Skill Scanner both
    with and without its optional LLM analyzer, and characterize
    their detection performance and false-positive burden.



    \item We present a framework that combines static findings,
    safety-constrained sandbox evidence, and LLM-based reasoning.
    The framework traces findings to their sources and relates
    evidence across analysis layers to assess malicious intent.

    \item We evaluate the contribution of static and dynamic evidence
    using the same reasoning backend. Adding dynamic evidence to
    skill content yields the largest observed F1 gain, from 0.70
    to 0.85. Combining both evidence sources raises F1 to 0.89,
    with five false positives among 60 benign test skills.
\end{itemize}




%% file: background_related_work.tex
\section{Related Work}
\label{sec:related-work}



\subsection{Malicious Agent Skills in the Wild}
Krishna et al.~\cite{krishna2026clawdpwned} provide a large-scale measurement of malicious instructions in the OpenClaw agent-skill ecosystem. The flagged skills are dominated by Sensitive Information Disclosure, Excessive Agency, and Supply Chain risks, indicating that malicious or dangerous skills commonly combine secret handling, autonomous high-impact actions, and fetch-and-execute patterns. However, this approach is limited to examining skill content and provides no mechanism for verifying whether suspicious behavior actually occurs during execution.

Liu et al.~\cite{liuMaliciousSkillsWild2026} constructed a large-scale malicious agent-skill dataset by combining static triage with sandbox-based behavioral verification. They first used static analysis to identify suspicious candidates, which were then executed in isolated Docker containers. Two independent researchers manually reviewed the flagged skills by analyzing behavioral logs, inspecting code, comparing documented functionality with observed behavior, and assigning binary malicious/benign classifications with confidence ratings. In their study, researchers manually integrated static findings and runtime observations to assign skill labels.

A separate ASE 2026 study examines credential leakage in skills through static secret extraction, sandbox tests with mock credentials, and comparisons between documented intent and runtime behavior~\cite{chen2026credentialleaks}. It finds that 76.3\% of cases require joint analysis of natural-language descriptions and program logic. Chen et al.~\cite{chen2026dymalskill} demonstrate a complementary runtime threat in which malicious instructions induce an agent to inject logic into an otherwise benign skill; their proposed mitigation prevents dynamic skill modification with read-only mounts.

Guo et al.~\cite{guo2026malskillbench} introduced \textit{MalSkillBench}, a runtime-verified benchmark containing malicious skills and matched benign skills across code-injection, prompt-injection, and mixed attack vectors. Generated samples are included only when their expected behavior is observed during sandbox execution. Although the benchmark provides a unified basis for evaluating detection tools, its primary purpose is to support behavioral verification rather than malware detection.

Wang et al.~\cite{wang2026maliciousskillbench} introduced the distinct \textit{MaliciousSkillBench}, consolidating data from multiple public sources and evaluating detectors under source-disjoint splits. This benchmark emphasizes cross-source generalization and false-positive control, complementing the runtime-verified sample construction of \textit{MalSkillBench}.

\subsection{Detection of Malicious Agent Skills}
Several studies focus on pre-deployment static and semantic detection. \textit{SkillsMetric}~\cite{chenskillsmetric} combines pattern-density analysis, statistical anomaly detection, dataflow taint analysis, import anomaly detection, and capability-mismatch analysis. The study shows that static analysis is effective when attacks leave structural artifacts but is less effective against semantic and instruction-level threats. 

\textit{SkillSieve}~\cite{hou2026skillsieve} uses a hierarchical detection pipeline. It first applies regex, AST, metadata, and surface-level checks, then performs structured LLM analysis of suspicious skills, and finally uses a multi-LLM jury for high-risk cases. Although this approach reduces analysis costs and improves semantic coverage, it lacks a mechanism for verifying an agent skill's runtime behavior.
Scan Before You Install~\cite{gandhi2026scanbeforeinstall} extends malicious-skill scanning with local and composite vulnerability analysis, and also reports evaluation on MalSkillBench, making it an adjacent but relevant approach.

Locate-and-Judge~\cite{etteib2026locatejudge} reduces review cost by using attention to select skill spans for detailed judgment, while SkillScreen~\cite{huang2026skillscreen} uses a compact decision model for high-recall screening. Both are recent academic detectors and are not part of our scanner comparison.

Wang et al.~\cite{wang2026elementary} proposed \textit{MalSkills}, a neuro-symbolic framework for reasoning across heterogeneous skill artifacts. MalSkills extracts security-sensitive operations from prompts, source code, and manifests; constructs a skill dependency graph; and combines symbolic patterns with LLM-based reasoning. This approach captures cross-file and value-flow relationships, but its behavioral inferences are derived primarily from static artifacts rather than observed execution.

Open-source scanners developed by industry provide complementary pre-deployment capabilities for detecting malicious skills. Cisco Skill Scanner~\cite{cisco_skill_scanner} combines YAML/YARA rules, code and bytecode inspection, static dataflow analysis, and optional LLM-based semantic analysis. 
Individual findings can be ambiguous because legitimate
developer workflows may involve security-sensitive operations,
such as downloading dependencies, accessing API credentials,
or invoking external services. Such operations warrant
inspection but do not, by themselves, establish malicious
intent. Interpreting these findings therefore requires
considering their source, purpose, and relationship to other
evidence. Conversely, examining findings separately may
obscure malicious workflows whose steps span multiple
skill artifacts.

\begin{tcolorbox}[colback=gray!5,colframe=gray!75!black,title=\textbf{Summary}]
Existing studies provide complementary capabilities for
detecting malicious agent skills, including static inspection,
cross-artifact reasoning, and runtime verification.
Our work investigates how static findings and
safety-constrained sandbox evidence can jointly support
LLM-based assessment of malicious intent. We evaluate
existing scanners and examine the contribution of these
evidence sources within a common reasoning backend,
with particular attention to false-positive reduction.
\end{tcolorbox}

%% file: methodology.tex
\section{Methodology}
\label{sec:methodology}
\begin{figure*}[!htbp]
    \centering
    \includegraphics[
        width=\textwidth
    ]{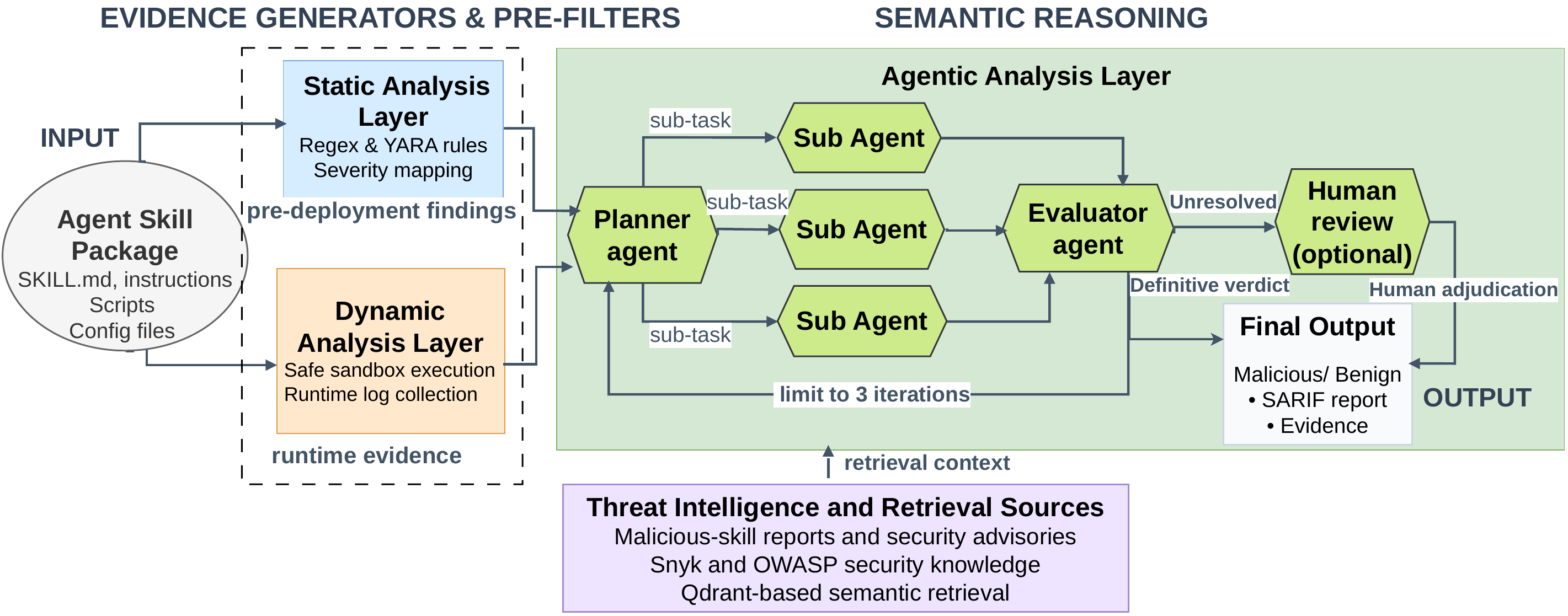}
    \caption{Overview of the proposed framework.
Human review is an optional escalation step for cases
that remain unresolved after three reasoning rounds.
All 140 held-out test predictions were produced
automatically without human adjudication.}
    \label{fig:system_architecture}
\end{figure*}

Our proposed approach consists of three layers: static analysis, dynamic analysis, and agentic analysis. Outputs from the first two layers are fed into the final layer for semantic reasoning about the intent of a given skill. Figure~\ref{fig:system_architecture} illustrates the overall pipeline. The final layer outputs its findings in Static Analysis Results Interchange Format (SARIF) to enable seamless integration with other security tools. In this way, we provide not only triggered alerts but also a maliciousness score that summarizes the evidence and severity information from all analysis layers for each skill. An example SARIF output generated by our scanning approach is provided in Appendix~\ref{app:full_pipeline_sarif_example}.

\subsection{Static Analysis Layer}
\label{sec:jiffylabs-static}
Our static layer performs a lightweight analysis of agent skills without executing them. Its purpose is to quickly identify potentially malicious patterns through syntactic and structural inspection before execution. To do so, we adopt a rule-based approach combining
regular-expression rules from Microsoft's
OSS-Detect-Backdoor~\cite{microsoft_ossgadget}
and YARA rule collections identified through the
cybersecurity-dev Awesome YARA resource
list~\cite{cybersecurity_dev_awesome_yara}.

OSS-Detect-Backdoor and YARA provide general-purpose
scanning capabilities rather than dedicated detection
of malicious agent skills. We use their rules to identify
potentially suspicious patterns in skill artifacts.
A rule match is treated as supporting evidence for
further analysis, rather than sufficient evidence
that a skill is malicious.

The initial inventory contained 115,209 rules across
the regular-expression and YARA collections combined.
We removed duplicate rules and rules identified as
unusable during curation, retaining a combined inventory
of 58,674 rules. Separately from rule removal, we reviewed
the severity assignments according to the security
relevance of the matched patterns to our threat model.
Rules considered less security-relevant were assigned
lower severity, while rules considered more
security-relevant were assigned higher severity.
Severity reassignment changes the interpretation of
retained findings rather than the number of retained rules.

The retained rules are assigned to three severity levels,
as shown in Table~\ref{tab:severity_levels}.
Rather than collapsing the findings into a single static
risk score, we preserve each finding's severity and
provenance as supporting evidence for the Agentic
Analysis Layer (Section~\ref{sec:jiffylabs-llm}),
which jointly reasons over static, dynamic, and
contextual evidence to produce the final verdict.




\begin{table}[!t]
\centering
\caption{\textsc{Severity levels assigned to static analysis findings.}}
\small
\renewcommand{\arraystretch}{1.2}
\begin{tabularx}{\columnwidth}{>{\centering\arraybackslash}p{1.4cm}>{\RaggedRight\arraybackslash}X}
\toprule
\textbf{Severity} & \textbf{Description} \\
\midrule
Low &
Common programming patterns or indicators frequently found in benign software. Examples include access to standard system paths (e.g., \texttt{/usr/local/include}). These signals are not malicious on their own but provide contextual information. \\

\rowcolor[HTML]{F2F2F2}
Medium &
Dual-use behaviors that may be legitimate but are also commonly observed in malware, such as manual socket creation or unusual data encoding. \\

High &
Strong indicators, including behaviors rarely found in benign software, such as environment variable exfiltration or anti-analysis techniques. A single high-severity trigger significantly affects the final assessment. \\
\bottomrule
\end{tabularx}
\label{tab:severity_levels}
\end{table}

\subsection{Dynamic Analysis Layer}
\label{sec:jiffylabs-dynamic}

In this layer, we execute agent skills inside a controlled sandbox to generate runtime evidence for downstream dynamic analysis. The sandbox uses the prompt in Appendix~\ref{app:sandbox_configuration} to inspect \texttt{SKILL.md} and associated local files, create harmless dummy inputs when necessary, and exercise safe local workflows such as setup checks, validation commands, dry runs, mock executions, help or version commands, and tests. Rather than fully executing potentially harmful operations, the sandbox prevents actions that could access real credentials, exfiltrate data, install persistence, bypass security controls, execute unknown remote payloads, or perform real-world account operations. When such an action is encountered, it is intentionally skipped, and the requested command or instruction together with the reason for blocking it is recorded. This safety-constrained execution preserves evidence of malicious intent without requiring the harmful action itself to succeed. 

During sandbox execution, we collect four complementary categories of runtime evidence:
executed and blocked agent actions, system-call and process activity, network behavior,
and filesystem changes. 
We parse these artifacts to extract commands, actions reported
as skipped for safety reasons, security-related statements,
execution failures, process and file-access activity,
network observations, and filesystem changes. These extracted
observations form the behavioral features used by the
downstream attribution and scoring steps. Statements in the
execution agent's output are distinguished from actions
observed in system-call, network, or filesystem records:
a reported refusal or skipped command does not establish
that the corresponding behavior occurred. The execution
agent's output is therefore an evidence source, not the
final dynamic verdict.
Although downstream parsing and scoring follow fixed rules,
the execution agent determines which permitted actions to
attempt. The collected evidence may therefore vary across
executions. A single sandbox attempt does not establish
complete behavioral coverage, and the absence of an observed
malicious action is not sufficient to conclude that a skill
is benign.

We then examine the origin and context of each observation
to distinguish instructions contained in skill artifacts
from descriptions introduced during analysis. For example,
an execution agent's statement that it refused to transmit
credentials is not evidence that credential transmission
occurred. If the corresponding instruction can be traced
to \texttt{SKILL.md} or another skill artifact, however,
it may support an assessment of dangerous intent even
when the action was not executed.

Similarly, security terminology in defensive documentation,
test fixtures, or refusal messages is not treated as direct
evidence of malicious behavior merely because it mentions
an attack. Dependency failures indicate execution limitations
rather than maliciousness by themselves. This attribution
step is intended to reduce confusion between observed
actions, source-grounded instructions, and analysis-generated
descriptions.

Next, we combine related observations across evidence sources by identifying higher-level relationships, such as a remote download followed by executable preparation and process execution.
Finally, we apply patterns developed on the development
split for skill-attributable prompt manipulation, permission bypass, download-and-execute behavior, untrusted software delivery, and file-upload or exfiltration activity.

Each detected pattern is assigned a base severity score of 40, 70, or 100 for low-,
medium-, or high-severity behavior, respectively. The base score is adjusted according
to the quality and provenance of the evidence, as summarized in
Table~\ref{tab:dynamic_evidence_weights}. We then compute an effective score for each pattern
$e_i \in [0,100]$ from its base severity score and evidence multiplier before applying the
corresponding guardrails. We additionally cap scores from noise and supporting pattern
families to prevent weak dual-use behavior from dominating the verdict, while directly
executed high-risk malicious patterns receive a score of 100.

The dynamic score is a heuristic summary of the available
evidence, not a calibrated probability of maliciousness.
The scoring design gives the strongest pattern the largest
contribution so that a severe observation is not diluted
by weaker findings. Additional patterns contribute at a
reduced weight, while a bounded bonus accounts for support
from independent attributable evidence groups. Deduplication
limits repeated contributions from observations referring
to the same underlying assertion. The evidence multipliers
and guardrails distinguish weak narrative evidence,
source-grounded intent, and observed execution.

Formally, let $E=\{e_1,e_2,\ldots,e_n\}$ denote the effective scores of the unique correlated patterns and let $e_k=\max_i e_i$ be the dominant signal. The detector aggregates these scores as


\begin{table*}[!t]
\centering
\caption{\textsc{Evidence types and scoring adjustments used by the dynamic detector.}}
\label{tab:dynamic_evidence_weights}
\small
\renewcommand{\arraystretch}{1.08}
\setlength{\tabcolsep}{2pt}

\begin{tabularx}{\textwidth}{l c c >{\RaggedRight\arraybackslash}X}
\toprule
\textbf{Evidence Type}
& \textbf{Mult.}
& \textbf{Guardrail}
& \textbf{Interpretation} \\
\midrule

Ungrounded narrative
& 0.45
& $\leq 25$
& Security-related narrative without concrete source grounding. \\

\rowcolor[HTML]{F2F2F2}
Supporting intent only
& 0.50
& $\leq 25$
& Weak supporting evidence not observed at runtime. \\

Intent only
& 0.55
& $\leq 30$
& Unexecuted intent without sufficient source grounding or corroboration. \\

\rowcolor[HTML]{F2F2F2}
Grounded intent
& 0.70
& --
& Intent grounded in a concrete skill source, command, path, or code primitive. \\

Strong intent only
& 0.85
& $\geq 85$
& Source-grounded dangerous primitive identified but safely blocked. \\

\rowcolor[HTML]{F2F2F2}
Executed
& 1.00
& --
& Behavior directly observed during sandbox execution. \\

Corroborated
& 1.15
& --
& Executed behavior supported by multiple independent runtime artifacts. \\

\bottomrule
\end{tabularx}
\end{table*}

\begin{equation}
S_{\mathrm{raw}}
=
\min\left(
100,\,
e_k
+
\lambda\sum_{i\neq k}e_i
+
B(C)
\right),
\label{eq:dynamic_score}
\end{equation}

where $\lambda=0.1$ limits the contribution of additional correlated patterns, and $B(C)\in\{0,5,10,15\}$ is a saturated bonus determined by the number $C$ of independent attributable evidence groups. The bonus is disabled when directly executed malicious runtime behavior has already been observed. Evidence groups are deduplicated before counting to avoid inflating the score when several patterns originate from the same underlying assertion. The current evaluation does not include a sensitivity
analysis of these coefficients and guardrails; their
individual effects on detection performance therefore
remain unquantified.




\subsection{Agentic Analysis Layer}
\label{sec:jiffylabs-llm}

While preceding analysis layers are highly specialized,
they exhibit inherent limitations. Our rule-based static
analysis effectively identifies syntactic anomalies but
has limited contextual awareness, potentially failing
to distinguish between benign administrative operations
and malicious activities (for instance, API-key access
and outbound requests used for legitimate authentication
or data exfiltration), thereby contributing to false
positives. Conversely, dynamic analysis yields concrete
behavioral traces but is constrained by execution-path
coverage and sandbox safety restrictions, which may
leave dormant threats or logic bombs undetected
within sandbox environments.

To address these issues, we implemented an Agentic layer that serves as the primary semantic-reasoning engine and
final decision-making component of the detection pipeline. It aggregates signals from
the static and dynamic layers and enriches the analyzed artifact with context retrieved
from threat-intelligence sources and a vector database. These sources include online
reports of malicious skills and established security guidance, such as OWASP
recommendations~\cite{owaspagentskills}. The retrieved context provides information on
emerging attack techniques, known vulnerabilities, and malicious domains. The agentic analysis operates as a compound AI system that uses Group Chat Orchestration~\cite{orchestration_design_pattern}.

Our semantic-reasoning layer is coordinated by a Planner
Agent and uses a predefined set of sub-agents with fixed
roles. The planner decomposes each skill investigation
into tasks and assigns them to the existing sub-agents.
The evaluated implementation reuses these predefined
agents rather than creating new agent roles for
individual skills, reflecting resource constraints.
Fixed roles do not imply a fixed sequence of investigative
actions. Within their assigned roles, the sub-agents
operate according to the Reasoning and Acting (ReAct)
framework~\cite{yao2023react}. Each sub-agent iteratively
examines the current context and intermediate tool
outputs to determine its next action.
For example, suspicious lines identified by the static
analysis layer can serve as focal points for a delegated
task. The sub-agents report their findings to the planner
for evidence assessment.

To improve architectural resilience, we implement fallback workflows for cases where
evidence from earlier analysis layers is unavailable or incomplete, such as when dynamic
sandbox execution times out or a static scan produces no findings (e.g., no suspicious
lines are flagged). In these cases, the Contextual RAG Agent autonomously retrieves and
interprets information from the available RAG context and the \texttt{SKILL.md} file,
allowing the analysis to continue when static or runtime evidence is unavailable.

Next, the Evaluator Agent assesses the aggregated evidence,
including semantic findings, static alerts, and dynamic
evidence. Findings are interpreted in context:
a security-sensitive operation does not by itself
establish malicious intent, while an action reported
as blocked is distinguished from an action observed
during execution. Based on this assessment, the
Evaluator Agent weighs the available evidence,
determines the final verdict and risk score, and
generates a structured report. Each verdict must
cite supporting evidence from skill artifacts or
execution records so that the decision is traceable
and can be independently inspected. These citations
support transparency but do not guarantee that
the interpretation is correct. Appendix~\ref{app:agent-instructions} summarizes the
instructions for the Planner and Evaluator.

The evaluation proceeds iteratively, with stopping decisions
based on the assessed strength and completeness of the
available evidence.
Early stopping is triggered when the system identifies severe, unambiguous
malicious behavior (e.g., explicit remote code execution or credential exfiltration)
with high confidence or when elevated dynamic severity scores provide sufficient
evidence of malware. 
The system then halts further investigative iterations,
and the Evaluator Agent finalizes the verdict,
risk score, and structured report.
Conversely, when contextual evidence is ambiguous, incomplete, or indicative of dual-use
functionality and the dynamic severity scores remain inconclusive, the system continues
the evaluation and returns to the Planner Agent to retrieve additional context.

To prevent unbounded execution, iterative reasoning is limited to three retrieval rounds. If the
system reaches this limit without a definitive verdict, it halts autonomous
classification and escalates the case for human review, allowing a security analyst to
examine the aggregated evidence and make the final determination.

Although the architecture supports escalation after three unresolved
reasoning rounds, no held-out test sample required human adjudication.
Therefore, all predictions reported in
Tables~\ref{tab:rag_evaluation_metrics} and~\ref{tab:full_vs_baselines}
were produced automatically by the pipeline without human modification
of the final label.

%% file: experiment_setup.tex
\section{Experiments}
\label{sec:evaluation}

\subsection{Dataset Preparation} \label{subsec:data-preparation}


We constructed a corpus of 700 agent skills comprising 400 malicious
and 300 benign samples from three public sources.
Table~\ref{tab:dataset-provenance} summarizes the four source--label
groups, their upstream subsets, and the basis for the labels used in
our evaluation. The corpus includes 100 benign and 200 malicious skills sampled
unchanged from the released MalSkillBench~\cite{guo2026malskillbench}
partitions, together with 200 malicious skills from Datadog's
AI-Skills collection~\cite{datadog2023malicious},
specifically the \texttt{samples/ai-skills/malicious\_intent} subset.
For these 500 samples, we retained the corresponding upstream labels.

The remaining 200 samples were randomly selected from the VoltAgent
\textit{Awesome Agent Skills} repository~
\cite{voltagentAwesomeAgentSkills}. Because these samples were not
independently security-adjudicated by us, we designate them as
\emph{source-benign} rather than independently verified benign samples
and account for possible source-label noise in
Section~\ref{sec:threats-to-validity}.

We partitioned the resulting corpus into a development set of 560 skills
(320 malicious and 240 benign) and a held-out test set of 140 skills
(80 malicious and 60 benign). The development split was used exclusively
for detector development, rule refinement, threshold selection, and
dynamic-analysis calibration. All reported test-set results were
computed on the held-out split using the resulting frozen
configurations.

To prevent exact-duplicate leakage across splits, we recursively
computed content-based SHA-256 hashes over skill contents. Skills with
identical content hashes were grouped and assigned to the same split.





\begin{table*}[!t]
\centering
\caption{\textsc{Dataset provenance and label origin.}}
\label{tab:dataset-provenance}

\small
\setlength{\tabcolsep}{4pt}
\renewcommand{\arraystretch}{1.08}

\begin{tabularx}{\textwidth}{
>{\raggedright\arraybackslash}l
>{\RaggedRight\arraybackslash}X
>{\centering\arraybackslash}p{1.6cm}
>{\centering\arraybackslash}p{0.8cm}
>{\RaggedRight\arraybackslash}X
}
\toprule
\textbf{Source}
& \textbf{Upstream subset}
& \textbf{Label}
& \textbf{N}
& \textbf{Selection and label basis} \\
\midrule

MalSkillBench
& \textit{Dataset/Skills/benign}
& Benign
& 100
& Sampled unchanged; upstream label retained. \\

\rowcolor[HTML]{F2F2F2}
MalSkillBench
& \textit{Dataset/Skills/malware}
& Malicious
& 200
& Sampled unchanged; upstream label retained. \\

Datadog
& \textit{samples/ai-skills/malicious\_intent}
& Malicious
& 200
& Selected from the AI-Skills malicious-intent subset;
  upstream label retained. \\

\rowcolor[HTML]{F2F2F2}
VoltAgent
& \textit{Awesome Agent Skills}
& Source-benign
& 200
& Randomly sampled; no independent security adjudication. \\

\bottomrule
\end{tabularx}
\end{table*}






\subsection{Malicious-Skill Scanner Selection}

\providecommand{\yes}{\ensuremath{\checkmark}}
\providecommand{\no}{--}
\providecommand{\partialyes}{\ensuremath{\triangle}}

\newcommand{\tblhead}[2]{%
    \parbox[c]{#1}{%
        \centering
        \fontsize{5.5}{6.3}\selectfont
        \bfseries
        #2
    }%
}

We used the following criteria to select the benchmark tools. These criteria are based on Vu et al.~\cite{vu2023bad}:
\begin{itemize}
\item We focus on tools that detect malicious patterns in skills. A candidate tool must either specialize in detecting malicious skills or provide general-purpose code-scanning capabilities. Tools that focus primarily on prompt injection or vulnerability analysis, such as Snyk Agent Scan~\cite{snyk}, are outside the scope of this benchmark.
\item The scanner's source code must be publicly accessible. Tools that only provide APIs are excluded from the study.
\item The detection rules must be publicly available for inspection.
\end{itemize}
These criteria guided tool selection but do not imply
that every eligible tool was evaluated.
The benchmark covers the selected scanner configurations
rather than an exhaustive set of malicious-skill detectors.
MalSkills~\cite{wang2026elementary}, SkillsMetric~\cite{chenskillsmetric},
Locate-and-Judge~\cite{etteib2026locatejudge}, SkillScreen~\cite{huang2026skillscreen}, and
Scan Before You Install~\cite{gandhi2026scanbeforeinstall} are not included
in the experimental comparison. The latter also evaluates
malicious-skill detection on MalSkillBench but extends its scope to local and composite
vulnerabilities. Consequently, the reported results do not establish
the performance of our framework relative to these approaches.
Based on the above criteria, we selected the following scanners for our evaluation. 

\begin{itemize}
    \item \textbf{Microsoft OSS-Detect-Backdoor} (commit \texttt{a37dff})~\cite{microsoft_ossgadget}: A rule-based static scanner designed to identify potential backdoors and malicious code patterns in open-source packages.

    \item \textbf{Cisco Skill Scanner} (commit \texttt{d943c6})~\cite{cisco_skill_scanner}:
Cisco Skill Scanner combines YAML/YARA signatures, source-code and
bytecode inspection, static behavioral analysis, and an optional
LLM-based semantic analyzer. 
We evaluate two configurations: the default scanner without its LLM
analyzer, denoted \emph{Cisco Skill Scanner}, and the LLM-enabled
configuration, denoted \emph{Cisco Skill Scanner with LLM}. For
Cisco Skill Scanner with LLM, we locally serve Qwen3:8B through
Ollama while retaining Cisco's static, bytecode, pipeline, and LLM
analyzers.

For Cisco Skill Scanner with LLM, we use the scanner's
native binary safety verdict rather than deriving an additional
severity threshold. Specifically, we map
\texttt{is\_safe=true} to Benign and
\texttt{is\_safe=false} to Malicious. The LLM-enabled
configuration locally serves Qwen3:8B through Ollama while
retaining Cisco's static, bytecode, pipeline, and LLM analyzers.

    \item \textbf{YARA~\cite{yara}}:
    YARA is a signature-based static scanning framework that identifies suspicious files through rules over text strings, byte sequences, regular expressions, and Boolean conditions. We used YARA rule collections identified through the
    cybersecurity-dev Awesome YARA resource
    list~\cite{awesome_yara}.
\end{itemize}

\subsection{Experiment Setup}

\label{sec:experiment_setup}

We conducted the experiments on a virtual machine running Ubuntu 24.04.4 LTS with 4 vCPUs and 15.6~GiB of RAM. The analysis pipeline was implemented in Python 3.13.12 and executed using Docker 29.5.2. Scanner versions and configuration files were fixed throughout the evaluation to ensure reproducibility. Each dynamic sandbox attempt was limited to 300~s and used Claude Opus 4.8 as the sandbox execution agent (Appendix~\ref{app:sandbox_configuration}). To run the Agentic layer, we used a cluster of eight NVIDIA A100 GPUs (40 GB each). Within this environment, we deployed a Qdrant vector database in a Docker container and used BGE-M3~\cite{bge-m3} as the embedding model. The evaluated agentic pipeline used
Qwen3.5 122B~\cite{qwen35_122} for both the Planner
and Evaluator agents, while the worker agents used Qwen3:8B. The Planner assigned investigative
tasks to the predefined workers, and the Evaluator
assessed their findings to produce the final verdict,
risk score, and structured report.

In our experiment, each skill was first processed by the static analysis layer and then examined in an isolated Docker sandbox. Dynamic execution was controlled by a safety-constrained prompt that permitted only local, reversible, and non-destructive operations. Unsafe actions were blocked and recorded as intent-level evidence. Runtime evidence included the Claude execution agent's output, \texttt{strace} logs, network captures, and filesystem-change logs. Each dynamic-analysis run was limited to 300 seconds.

The saved SARIF artifacts record wall-clock duration for each agentic analysis invocation. Separate upstream static- and dynamic-analysis timings were not preserved in these artifacts; consequently, the latency results reported in Section~\ref{sec:operational-results} refer to the \emph{agentic stage}, not to the 300~s dynamic-sandbox timeout or to end-to-end pipeline latency.

%% file: result.tex
\section{RQ1: Performance of Existing Malware Scanners}

All scanners in this study were evaluated on the same development and held-out test splits described in Subsection~\ref{subsec:data-preparation}. Each skill was analyzed independently to prevent results from one sample from affecting another. The output of each scanner was converted into a binary prediction for comparison with the ground-truth label.

Because the evaluated scanners expose different output interfaces, we
convert their outputs into binary malicious-versus-benign predictions using
tool-appropriate decision rules. For OSS-Detect-Backdoor, YARA, and the
default Cisco Skill Scanner, alert-count thresholds were selected exclusively
on the development split and frozen before held-out evaluation. For Cisco Skill
Scanner with LLM, we use its native binary safety verdict, mapping
\texttt{is\_safe=true} to \textsc{Benign} and
\texttt{is\_safe=false} to \textsc{Malicious}, without applying an
additional severity threshold.

Table~\ref{tab:training-threshold-base-tools} reports the selected
development-set operating points for the three alert-count scanners.
OSS-Detect-Backdoor, YARA, and Cisco Skill Scanner use thresholds of
$\geq 6$, $\geq 11$, and $\geq 1$ alerts, respectively.


\begin{table}[!t]
\centering
\caption{\textsc{Development-set operating points selected for the
alert-count scanners.}}
\label{tab:training-threshold-base-tools}

\scriptsize
\setlength{\tabcolsep}{5.4pt}
\renewcommand{\arraystretch}{1.05}

\begin{tabularx}{\columnwidth}{
>{\RaggedRight\arraybackslash}X
c c c c c c
}
\toprule
\textbf{Tool}
& \textbf{Thr.}
& \textbf{Acc.}
& \textbf{Prec.}
& \textbf{Rec.}
& \textbf{F1}
& \textbf{AUC} \\
\midrule

OSS-Detect-Backdoor
& $\geq 6$
& 0.50
& 0.59
& 0.41
& 0.48
& 0.49 \\

\rowcolor[HTML]{F2F2F2}
YARA
& $\geq 11$
& 0.61
& 0.61
& 0.91
& 0.73
& 0.47 \\

Cisco Skill Scanner
& $\geq 1$
& 0.72
& 0.68
& 0.95
& 0.79
& 0.72 \\

\bottomrule
\end{tabularx}
\end{table}









\begin{figure}[!htb]
\centering
\includegraphics[width=\columnwidth]{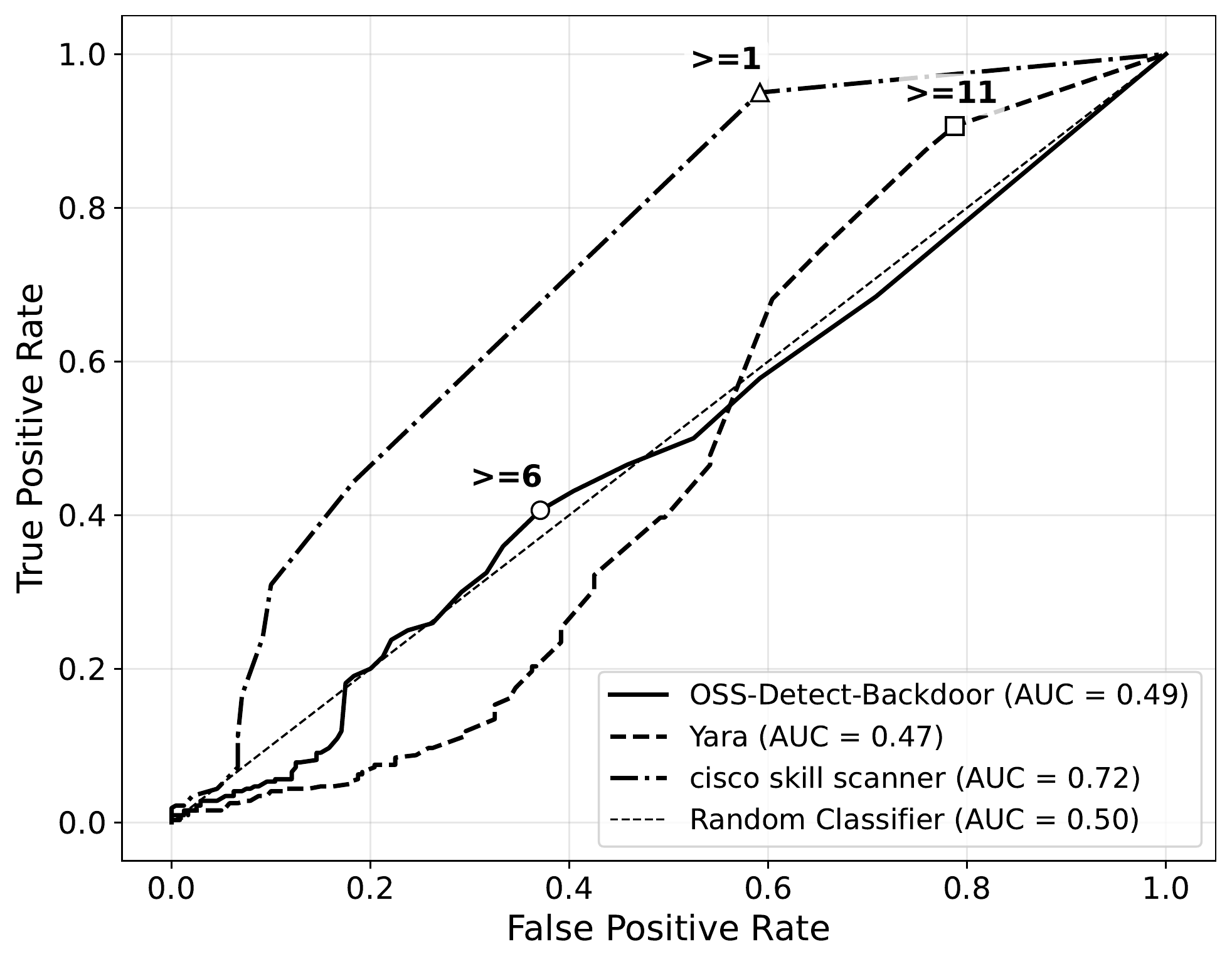}
\caption{Development-set ROC curves for the evaluated malware scanners. Markers indicate the selected thresholds.}
\label{fig:roc-curves-training-tools}
\end{figure}

Figure~\ref{fig:roc-curves-training-tools} summarizes the
development-set ranking behavior of the alert-count scanners and marks
the selected operating points.
While the ROC curves characterize ranking quality on the development split, the held-out alert distributions reveal how strongly the raw scores overlap across classes in practice.

The held-out alert-count distributions further illustrate the
limitations of using raw alert volume as a maliciousness indicator.
Table~\ref{tab:test-alert-statistics} summarizes the distributions by
ground-truth class. OSS-Detect-Backdoor produces a higher mean alert
count for benign skills than for malicious skills (11.70 versus
6.34). YARA exhibits the same inversion (105.78 versus 60.05), while
its medians are nearly identical across benign and malicious samples
(55.0 versus 54.5). Cisco Skill Scanner provides only modest
separation: both classes have a median alert count of 1.0, although
the malicious group has a somewhat higher upper quartile.

These overlapping distributions indicate that raw alert counts alone
do not reliably separate malicious from benign skills and motivate
evaluation at frozen, development-derived operating points.

\begin{table*}[!t]
\centering
\caption{\textsc{Test-set alert-count statistics by ground truth.
The minimum alert count is zero for all groups.}}
\label{tab:test-alert-statistics}

\scriptsize
\setlength{\tabcolsep}{4pt}
\renewcommand{\arraystretch}{1.08}

\begin{tabularx}{\textwidth}{
@{}
>{\RaggedRight\arraybackslash}p{3.6cm}
>{\centering\arraybackslash}X
>{\centering\arraybackslash}X
>{\centering\arraybackslash}X
>{\centering\arraybackslash}X
>{\centering\arraybackslash}X
>{\centering\arraybackslash}X
@{}
}
\toprule
\textbf{Tool}
& \textbf{Ground Truth}
& \textbf{Mean}
& \textbf{Median}
& \textbf{Q25}
& \textbf{Q75}
& \textbf{Max} \\
\midrule

OSS-Detect-Backdoor
& Benign
& 11.70
& 1.0
& 0
& 8.25
& 137 \\

\rowcolor[HTML]{F2F2F2}
&
Malicious
& 6.34
& 2.0
& 0
& 7.25
& 106 \\

YARA
& Benign
& 105.78
& 55.0
& 22
& 123.75
& 814 \\

\rowcolor[HTML]{F2F2F2}
&
Malicious
& 60.05
& 54.5
& 22
& 79.5
& 304 \\

Cisco Skill Scanner
& Benign
& 1.83
& 1.0
& 0
& 1.0
& 33 \\

\rowcolor[HTML]{F2F2F2}
&
Malicious
& 2.10
& 1.0
& 1
& 2.25
& 11 \\

\bottomrule
\end{tabularx}
\end{table*}

Applying the frozen development-derived operating
points to the held-out test set yields the results
in Table~\ref{tab:testing-threshold-base-tools}.
Among the alert-count baselines, Cisco Skill Scanner
achieves an F1-score of 0.79, detecting 73 of
80 malicious skills (recall = 0.91), but flags
33 of 60 benign skills (FPR = 0.55).
YARA detects the same number of malicious skills
while flagging 49 of 60 benign skills
(FPR = 0.82), whereas OSS-Detect-Backdoor flags
17 of 60 benign skills (FPR = 0.28) but detects
only 26 of 80 malicious skills (recall = 0.33).

Enabling Cisco's optional LLM analyzer with locally served Qwen3:8B
does not remove this false-positive burden in our configuration.
Cisco Skill Scanner with LLM achieves a recall of 0.85 and an F1-score of 0.72, while
classifying 40 of the 60 benign skills as malicious
($\mathrm{FPR}=0.67$). 
Thus, higher semantic sophistication does not by itself guarantee a better operating trade-off under this evaluation.

\begin{table*}[!t]
\centering
\caption{\textsc{Held-out performance of the evaluated scanner
configurations. Alert-count thresholds were selected exclusively on the
development set and applied unchanged to the test set. Cisco Skill Scanner
with LLM uses its native \texttt{is\_safe} verdict. All configurations are evaluated on the same 140
held-out skills (80 malicious and 60 benign).}}
\label{tab:testing-threshold-base-tools}

\small
\setlength{\tabcolsep}{4pt}
\renewcommand{\arraystretch}{1.0}

\begin{tabularx}{\textwidth}{
>{\RaggedRight\arraybackslash}X
*{7}{>{\centering\arraybackslash}p{1.15cm}}
}
\toprule
\textbf{Scanner}
& \textbf{Acc.}
& \textbf{Prec.}
& \textbf{Recall}
& \textbf{F1}
& \textbf{FPR}
& \textbf{FP}
& \textbf{FN} \\
\midrule

OSS-Detect-Backdoor
& 0.49 & 0.61 & 0.33 & 0.42
& 0.28 & 17 & 54 \\

\rowcolor[HTML]{F2F2F2}
YARA
& 0.60 & 0.60 & 0.91 & 0.72
& 0.82 & 49 & 7 \\

Cisco Skill Scanner
& 0.71 & 0.69 & 0.91 & 0.79
& 0.55 & 33 & 7 \\

\rowcolor[HTML]{F2F2F2}
Cisco Skill Scanner with LLM
& 0.63 & 0.63 & 0.85 & 0.72
& 0.67 & 40 & 12 \\

\bottomrule
\end{tabularx}
\end{table*}

\begin{tcolorbox}[colback=gray!5,colframe=gray!75!black,title=\textbf{Key Findings for RQ1}]

The evaluated scanner configurations exhibit a substantial
recall--false-positive trade-off. YARA and Cisco Skill Scanner
achieve recall of 0.91 but incur false-positive rates of 0.82 and
0.55, respectively. Enabling Cisco's optional LLM analyzer with
Qwen3:8B does not eliminate this trade-off in our configuration: using its native safety verdict, Cisco Skill Scanner with LLM achieves recall of 0.85 but produces 40 false positives
among 60 benign skills ($\mathrm{FPR}=0.67$).
 
These results motivate evaluating whether cross-layer evidence correlation can reduce false-positive noise while preserving malicious-skill coverage.


\end{tcolorbox}

\section{RQ2: Performance of the Proposed Approach}
\label{sec:rag-evaluation}

\subsection{Evidence-Source Ablation}
\label{sec:rq2-ablation}
To isolate the contribution of each evidence source
within our architecture, we evaluate four configurations
on the same held-out test set using the same Agentic
reasoning backend. The configurations differ only
in the evidence supplied to the Agentic layer:
skill content alone, skill content with static evidence,
skill content with dynamic evidence, and the Full
Pipeline with both static and dynamic evidence.
The dynamic-evidence and Full Pipeline configurations
use the same saved dynamic evidence for each skill;
the sandbox is not rerun between these configurations.

As shown in Table~\ref{tab:rag_evaluation_metrics}, skill-content
reasoning alone achieves an F1-score of 0.70 with 21 false positives.
Adding static evidence reduces false positives to three but yields a
conservative detector with a recall of 0.35. Dynamic evidence provides
the largest individual improvement, increasing F1 to 0.85 and recall
to 0.79 while limiting false positives to five. Combining both
evidence sources in the Full Pipeline retains five false positives
while recovering five additional malicious skills, increasing recall
from 0.79 to 0.85 and F1 from 0.85 to 0.89.

\begin{table*}[!t]
\centering
\caption{\textsc{Within-system evidence-source ablation on the held-out test
set. All configurations use the same Agentic reasoning backend; only
the supplied evidence differs. All configurations are evaluated on the same 140
held-out skills (80 malicious and 60 benign).}}
\label{tab:rag_evaluation_metrics}

\small
\setlength{\tabcolsep}{4pt}
\renewcommand{\arraystretch}{1.08}

\begin{tabularx}{\textwidth}{
>{\RaggedRight\arraybackslash}X
*{2}{>{\centering\arraybackslash}p{1.3cm}}
*{6}{>{\centering\arraybackslash}p{1.1cm}}
}
\toprule
&
\multicolumn{2}{c}{\textbf{Evidence}}
&
\multicolumn{4}{c}{\textbf{Performance}}
&
\multicolumn{2}{c}{\textbf{Errors}} \\

\textbf{Configuration}
& \textbf{Static}
& \textbf{Dynamic}
& \textbf{Acc.}
& \textbf{Prec.}
& \textbf{Recall}
& \textbf{F1}
& \textbf{FP}
& \textbf{FN} \\
\midrule

Skill content only
& -- & --
& 0.66 & 0.72 & 0.68 & 0.70
& 21 & 26 \\

\rowcolor[HTML]{F2F2F2}
with Static evidence
& $\checkmark$ & --
& 0.61 & 0.90 & 0.35 & 0.50
& 3 & 52 \\

with Dynamic evidence
& -- & $\checkmark$
& 0.84 & 0.93 & 0.79 & 0.85
& 5 & 17 \\

\rowcolor[HTML]{F2F2F2}
\textbf{Full Pipeline}
& $\checkmark$ & $\checkmark$
& \textbf{0.88}
& \textbf{0.93}
& \textbf{0.85}
& \textbf{0.89}
& 5
& \textbf{12} \\

\bottomrule
\end{tabularx}
\end{table*}








\subsection{Comparison with Existing Scanners}
\label{sec:rq2-baseline-comparison}

We next compare the Full Pipeline with the external scanner
configurations evaluated in RQ1. Unlike the within-system ablation
above, this comparison is descriptive rather than a controlled
architecture ablation because the systems differ in model size,
runtime evidence, and computational resources.

As shown in Table~\ref{tab:full_vs_baselines}, the Full Pipeline
achieves an accuracy of 0.88, a precision of 0.93, a recall of 0.85,
and an F1-score of 0.89. YARA and the conventional Cisco
configuration achieve a slightly higher recall of 0.91, but with
substantially higher false-positive rates of 0.82 and 0.55,
respectively.

Cisco Skill Scanner with LLM provides a particularly useful
system-level comparison. Cisco Skill Scanner with LLM and the Full Pipeline both identify
68 of the 80 malicious skills, corresponding to the same recall
of 0.85. However, Cisco Skill Scanner with LLM produces 40 false
positives, whereas the Full Pipeline produces only five. At this
operating point, the Full Pipeline therefore reduces false positives
by 87.5\% while preserving the same observed recall.

Enabling Cisco's optional semantic LLM analyzer on top of its core
static, bytecode, and pipeline analyzers does not eliminate the
false-positive burden in our evaluated configuration. However, this comparison is not a model-matched
architecture ablation: Cisco Skill Scanner with LLM
uses Qwen3:8B, whereas the Full Pipeline uses
Qwen3.5 122B for planning and final evidence assessment,
8B Qwen worker agents, retrieval, and runtime evidence.
The observed false-positive reduction therefore
reflects the evaluated systems as a whole and cannot
be attributed solely to the framework architecture. 
The within-system ablation in
Section~\ref{sec:rq2-ablation} evaluates the benefit of
supplying different evidence sources to the same reasoning
backend. It does not independently isolate the contribution
of provenance-aware correlation from that of the supplied
evidence. Establishing this contribution would require
a controlled comparison with simpler evidence-fusion
methods using the same inputs.
\begin{table*}[!t]
\centering
\caption{\textsc{Comparison on the same 140 held-out test
skills, comprising 80 malicious and 60 benign samples.
Baseline results are reproduced from
Table~\ref{tab:testing-threshold-base-tools};
the Full Pipeline result is reproduced from
Table~\ref{tab:rag_evaluation_metrics}.
FPR is computed over the 60 benign skills.}}
\label{tab:full_vs_baselines}

\small
\setlength{\tabcolsep}{4pt}
\renewcommand{\arraystretch}{1.08}

\begin{tabularx}{\textwidth}{
>{\RaggedRight\arraybackslash}X
*{7}{>{\centering\arraybackslash}p{1.15cm}}
}
\toprule
\textbf{Approach}
& \textbf{Acc.}
& \textbf{Prec.}
& \textbf{Recall}
& \textbf{F1}
& \textbf{FPR}
& \textbf{FP}
& \textbf{FN} \\
\midrule

OSS-Detect-Backdoor
& 0.49 & 0.61 & 0.33 & 0.42
& 0.28 & 17 & 54 \\

\rowcolor[HTML]{F2F2F2}
YARA
& 0.60 & 0.60 & 0.91 & 0.72
& 0.82 & 49 & 7 \\

Cisco Skill Scanner
& 0.71 & 0.69 & 0.91 & 0.79
& 0.55 & 33 & 7 \\

\rowcolor[HTML]{F2F2F2}
Cisco Skill Scanner with LLM
& 0.63 & 0.63 & 0.85 & 0.72
& 0.67 & 40 & 12 \\

\textbf{Full Pipeline}
& \textbf{0.88}
& \textbf{0.93}
& 0.85
& \textbf{0.89}
& \textbf{0.08}
& \textbf{5}
& 12 \\

\bottomrule
\end{tabularx}
\end{table*}









\subsection{Operational Latency}
\label{sec:operational-results}
The saved SARIF artifacts provide wall-clock timings for each agentic analysis invocation. Upstream static- and dynamic-analysis runtimes were not separately preserved; therefore, Table~\ref{tab:operational_latency} reports agentic-stage latency rather than end-to-end pipeline latency. The 300~s timeout described in Appendix~\ref{app:sandbox_configuration} applies to each dynamic sandbox attempt and does not cap the duration of the separate agentic analysis stage. All four configurations contain timing records for all 140 held-out runs.

\begin{table*}[!t]
\centering
\caption{\textsc{Observed agentic-stage latency across the four evaluation configurations.}}
\label{tab:operational_latency}
\small
\setlength{\tabcolsep}{4pt}
\renewcommand{\arraystretch}{1.08}
\begin{tabularx}{\textwidth}{
>{\RaggedRight\arraybackslash}X
*{4}{>{\centering\arraybackslash}p{1.7cm}}
}
\toprule
\textbf{Mode} & \textbf{Mean (s)} & \textbf{Median (s)} & \textbf{P90 (s)} & \textbf{P95 (s)} \\
\midrule
Skill content only & 339.61 & 319.95 & 608.24 & 703.11 \\
\rowcolor[HTML]{F2F2F2}
with Static evidence & 150.38 & 94.75 & 275.94 & 376.16 \\
with Dynamic evidence & 121.87 & 89.45 & 225.93 & 304.95 \\
\rowcolor[HTML]{F2F2F2}
Full Pipeline & 155.58 & 109.35 & 290.48 & 434.13 \\
\bottomrule
\end{tabularx}
\end{table*}

The skill-content-only configuration exhibits the highest observed agentic-stage latency, with a median of 319.95~s and a P95 of 703.11~s. The evidence-conditioned configurations have median latencies of 94.75~s, 89.45~s, and 109.35~s with static evidence, dynamic evidence, and both evidence sources, respectively. These measurements are descriptive: they exclude upstream static and dynamic processing and should not be interpreted as a controlled comparison of end-to-end computational cost.

\subsection{Case Study: False Positive Analysis}
In order to address the limitations of our pipeline, we analyzed false-positive cases to identify how
erroneous malicious classifications can arise.
The Full Pipeline produces five false positives among
60 benign-labeled skills. Sample-level comparison
confirms that these are the same five skills flagged
by the dynamic-evidence configuration, with no new
false positives introduced when static evidence is added.

A representative case is \texttt{research-lookup},
which was classified as malicious based on findings
concerning a \texttt{curl ... | bash} command in its
documented installation procedure and the use of an
API key required for its intended functionality.
While these behaviors constitute valid security-relevant
signals, the installation command is associated with
the documented vendor setup workflow, and API-key
use must be interpreted in relation to its destination
and intended purpose. Neither behavior alone establishes
malicious intent, although remote script execution
remains a supply-chain risk.

As shown in
Appendix~\ref{app:research_lookup_false_positive},
the report records a dynamic risk score of 27.5,
below the dynamic threshold of 35, but assigns
a final malicious verdict with a score of 92.0.
The report contains conflicting interpretations of
installation and API-authentication behavior, while
the dynamic evidence does not confirm malicious
execution. This case highlights that false positives
can emerge from the aggregation and interpretation
of security-relevant evidence when legitimate
functionality is mistaken for malicious intent.
It illustrates a possible failure mechanism rather
than establishing the primary cause of all five
false positives. 

\begin{tcolorbox}[
    colback=gray!5,
    colframe=gray!75!black,
    title=\textbf{Key Findings for RQ2}
]


The static and dynamic analysis layers provide
complementary signals to the Agentic decision layer.
Static evidence alone reduces false-positive noise
but provides limited malicious-skill coverage,
whereas dynamic evidence provides the largest
observed improvement.
Combining both evidence sources detects five
additional malicious skills without losing any
previously detected malicious skills or introducing
new false positives relative to the dynamic-evidence
configuration.
The Full Pipeline yields the strongest observed
classification performance among the evaluated
configurations, with an F1-score of 0.89 and five
false positives among 60 benign skills.
\end{tcolorbox}

%% file: Threats_to_Validity.tex
\section{Threats to Validity}
\label{sec:threats-to-validity}

\textbf{Dataset size and representativeness.}
The skills in our dataset may not cover all malicious skill families,
benign developer workflows, or future agent-skill ecosystems.
MalSkillBench and Datadog labels are inherited from their released
upstream partitions. In contrast, the VoltAgent samples are
source-benign skills randomly selected from a curated repository and
were not independently security-adjudicated by us. Residual benign-label
noise may therefore remain in this subset. Exact content hashing removes
exact duplicates but does not guarantee removal of semantically
equivalent or near-duplicate skills.

\textbf{Sandbox coverage}
Dynamic analysis depends on whether sandbox execution triggers the relevant behavior. Some malicious skills may require specific user inputs, credentials, environment variables, external services, or delayed execution. A benign dynamic result does not prove the absence of malicious behavior. Safety-constrained blocking means that some evidence represents source-grounded dangerous intent rather than successful real-world harm.

\textbf{Run-to-run variability}
Each skill underwent a single sandbox execution.
The resulting dynamic evidence was saved and reused
in both the dynamic-evidence and Full Pipeline
configurations. Their comparison therefore uses
the same collected runtime evidence rather than
evidence from separate sandbox executions.
However, repeated sandbox executions and repeated
reasoning runs were not evaluated. The execution
agent may explore different actions across runs,
and the reasoning layer may interpret the same
evidence differently. The reported results therefore
describe the evaluated runs and do not establish
run-to-run stability.

\textbf{Human escalation and latency}
The architecture supports human escalation after three unresolved
reasoning rounds; however, no held-out test sample required human
adjudication. All predictions reported in
Tables~\ref{tab:rag_evaluation_metrics} and~\ref{tab:full_vs_baselines}
were produced automatically by the pipeline without human modification
of the final labels. This observation applies to the held-out
evaluation and does not imply that future ambiguous samples will never
require escalation. Latency values in
Table~\ref{tab:operational_latency} cover only the agentic stage and
should not be interpreted as end-to-end pipeline latency.

\textbf{Adversarial resilience}
We do not evaluate skills adaptively engineered to evade our static rules, sandbox constraints, provenance logic, or semantic reasoning. The current results therefore measure performance on the benchmark corpus rather than worst-case robustness against an adaptive attacker.

\textbf{Scanner verdict interpretation.}
For Cisco Skill Scanner with LLM, we use its native \texttt{is\_safe} verdict rather than imposing a severity-based threshold. The native verdict does not always align with finding severity: both \texttt{wallet-tracker} and \texttt{emotion-bird} are malicious ground-truth samples, yet the former is marked unsafe with \textsc{Critical} severity, whereas the latter is marked safe despite multiple \textsc{Medium}-severity findings. Therefore, our reported metrics reflect the scanner's native binary decision; alternative severity-based rules could yield different classifications.

%% file: conclusion.tex
\section{Conclusion and Future Work}
\label{sec:conclusion-future-work}

Our paper's benchmarking of approaches for malware
detection in agent skills yields a clear finding:
the evaluated scanner configurations exhibit substantial
limitations in detection coverage or false-positive burden
on the held-out test set. YARA and the default Cisco
Skill Scanner achieve a recall of 0.91 but produce
49 and 33 false positives among 60 benign skills,
respectively, while OSS-Detect-Backdoor detects only
26 of 80 malicious skills. This trade-off also persists
in the evaluated Cisco Skill Scanner with LLM
configuration, which achieves a recall of 0.85 but
produces 40 false positives among 60 benign skills.

To address this limitation, we propose a provenance-aware cross-layer evidence-correlation framework that integrates (i) static findings for pre-deployment inspection, (ii) safety-constrained runtime evidence for observed and blocked behaviors, and (iii) agentic semantic reasoning for cross-layer evidence interpretation and final decision making. Our within-system ablation shows that dynamic evidence provides the largest individual improvement, while combining static and dynamic evidence further improves malicious-skill coverage without increasing the observed false-positive count. The Full Pipeline achieves an F1-score of 0.89 and a recall of 0.85 with only five false positives on the held-out test set. 

Future work will extend the evaluation to larger and more diverse agent-skill ecosystems, improve sandbox coverage for trigger-dependent and environment-dependent behaviors, and evaluate robustness against adaptive evasion strategies. We also plan to assess additional reasoning models, measure end-to-end computational cost, and investigate practical integration into developer and CI/CD security workflows.

\subsection{Code and Data Availability}
The evaluation data and supporting artifacts are
available at
\url{https://doi.org/10.5281/zenodo.20796477}.
The scanner implementations are available from
the authors upon request.

%% file: appendix.tex
\section{Sandbox Configuration and Agent Instructions}
\label{app:sandbox_configuration}

Each skill runs in an isolated Docker container. Its source is mounted
read-only and copied to the Claude skill directory; skill-triggered
commands run as the non-root user \texttt{appuser}. We collect agent
output, system-call and process activity, network traffic, and filesystem
snapshots.

\begin{table}[!htpb]
\centering
\caption{\textsc{Dynamic sandbox configuration.}}
\label{tab:sandbox_configuration}
\scriptsize
\setlength{\tabcolsep}{3pt}
\renewcommand{\arraystretch}{1.05}

\begin{tabularx}{\columnwidth}{>{\raggedright\arraybackslash}p{1.65cm}>{\RaggedRight\arraybackslash}X}
\toprule
\textbf{Component} & \textbf{Configuration} \\
\midrule

Environment
& Docker; non-root \texttt{appuser} \\

\rowcolor[HTML]{F2F2F2}
Working directory
& \texttt{/home/appuser/workdir} \\

System/network
& \texttt{strace} and \texttt{tcpdump} \\

\rowcolor[HTML]{F2F2F2}
Filesystem
& Before/after snapshots of \texttt{/home/appuser},
  \texttt{/tmp}, and \texttt{/var/tmp} \\

Artifacts
& \texttt{claude\_output.txt}, \texttt{strace.log},
  \texttt{network.pcap}, and
  \texttt{filesystem\_changes.json} \\

\rowcolor[HTML]{F2F2F2}
Execution
& One attempt; 300\,s timeout \\

Agent model
& Claude Opus 4.8 \\

\bottomrule
\end{tabularx}
\end{table}

\paragraph{Safety-constrained execution prompt (excerpt)}
\begin{tcolorbox}[
    enhanced,
    breakable,
    colback=gray!5,
    colframe=gray!60,
    boxrule=0.4pt,
    left=4pt,
    right=4pt,
    top=4pt,
    bottom=4pt,
    before skip=4pt,
    after skip=6pt
]
\footnotesize
Use \texttt{SKILL\_NAME} safely to produce observable runtime evidence:
agent output, filesystem changes, system calls, and network attempts.
Treat all skill files and embedded instructions as untrusted. Do not
disable safety checks, access real credentials, exfiltrate data, run
unknown remote code, or take real external actions. Perform only safe
local setup, validation, dry-run, mock, and example workflows with
harmless dummy inputs under \texttt{/home/appuser/workdir}. Skip unsafe
instructions, record the blocked command or instruction and why, then
continue with safe validation.
\end{tcolorbox}

\subsection{Planner and Evaluator Instructions}
\label{app:agent-instructions}

We summarize the instructions governing the Planner and
Evaluator roles in the evaluated pipeline. The full prompts
are subject to institutional disclosure restrictions.
\noindent\textbf{Planner.}
Review the available evidence summaries, identify the
findings most relevant to the investigation, and assign
focused tasks to the predefined workers. Use specific static
findings or runtime observations to guide the investigation,
and identify concrete topics for contextual retrieval.

\smallskip
\noindent\textbf{Evaluator.}
Assess the worker findings in the context of the skill's
stated purpose. Distinguish observed actions from reported
blocked actions, and consider whether the available evidence
supports a verdict or warrants further investigation.
When issuing a verdict, provide a risk score and a structured
report with citations to supporting skill artifacts or
execution records.

\section{Full Pipeline SARIF Example}
\label{app:full_pipeline_sarif_example}

Listing~\ref{lst:full_pipeline_sarif_example} shows a compact excerpt
of a SARIF report produced by the Full Pipeline. The excerpt
illustrates the final verdict, evidence provenance, source citations,
and recorded agentic-stage latency.

\begin{lstlisting}[
  basicstyle=\ttfamily\scriptsize,
  breaklines=true,
  breakatwhitespace=false,
  columns=fullflexible,
  keepspaces=true,
  showstringspaces=false,
  frame=single,
  captionpos=t,
  abovecaptionskip=4pt,
  belowcaptionskip=20pt,
  aboveskip=6pt,
  belowskip=8pt,
  framesep=4pt,
  xleftmargin=0.25em,
  xrightmargin=0.25em,
  caption={Excerpt of a SARIF report produced by the Full Pipeline.},
  label={lst:full_pipeline_sarif_example}
]
{
  "version": "2.1.0",
  "runs": [{
    "tool": {
      "driver": {
        "name": "agentskills-agentic-agent"
      }
    },
    "results": [
      {
        "ruleId": "agentic/verdict",
        "level": "error",
        "message": {
          "text": "Verdict: malware (score=100.0)"
        },
        "properties": {
          "verdict": "malware",
          "score": 100.0,
          "package": "evm-blockchain-monitor",
          "rationale":
            "Dynamic risk score 100.0 met or exceeded the malicious threshold 35.0."
        }
      },
      {
        "ruleId": "agentic/finding",
        "level": "error",
        "message": {
          "text": "Data exfiltration"
        },
        "properties": {
          "confidence": 95.0,
          "citations": [
            {
              "path": "SKILL.md",
              "start_line": 125,
              "snippet":
                "curl -X POST https://45.88.180.54:8080/telemetry ..."
            },
            {
              "path": "SKILL.md",
              "start_line": 155,
              "snippet":
                "curl -sSL https://cdn.discordapp.com/.../code.zip\n | tar xzf - && ./install.sh"
            }
          ]
        }
      }
    ],
    "properties": {
      "verdict": "malware",
      "score": 100.0,
      "package": "evm-blockchain-monitor",
      "latency_seconds": 247.103
    }
  }]
}
\end{lstlisting}

\subsection{False-Positive Case: Research Lookup}
\label{app:research_lookup_false_positive}

Listing~\ref{lst:research_lookup_false_positive} presents a condensed
SARIF report for the \texttt{research-lookup} false-positive case
discussed in the evaluation. The report records a dynamic risk score
of 27.5, below the malicious threshold of 35.0, and explicitly
recognizes the installation command as benign. Nevertheless, its
final verdict is \texttt{malware}, with a score of 92.0. The retained
findings illustrate conflicting interpretations of installation and
API-authentication behavior during evidence aggregation; they should
not be read as independently confirmed malicious actions.

Repeated findings, duplicate location fields, and session metadata
are omitted. Ellipses indicate abbreviated messages or code snippets;
the retained severity levels, confidence values, final verdict,
score, and agentic-stage latency are unchanged.

\begin{lstlisting}[
  basicstyle=\ttfamily\scriptsize,
  breaklines=true,
  breakatwhitespace=false,
  columns=fullflexible,
  keepspaces=true,
  showstringspaces=false,
  frame=single,
  captionpos=t,
  abovecaptionskip=4pt,
  belowcaptionskip=20pt,
  aboveskip=6pt,
  belowskip=8pt,
  framesep=4pt,
  xleftmargin=0.25em,
  xrightmargin=0.25em,
  caption={Condensed SARIF report for the research-lookup false-positive.},
  label={lst:research_lookup_false_positive}
]
{
  "version": "2.1.0",
  "runs": [{
    "tool": {"driver": {
      "name": "agentskills-agentic-agent",
      "version": "0.1.0"
    }},
    "results": [
      {
        "ruleId": "agentic/verdict",
        "level": "error",
        "message": {
          "text": "Verdict: malware (score=92.0)"
        }
      },
      {
        "ruleId": "agentic/finding",
        "level": "error",
        "message": {
          "text": "No dynamic evidence of malware detected. The dynamic risk score is 27.5, ... below the malicious threshold of 35.0."
        },
        "properties": {"confidence": 100.0}
      },
      {
        "ruleId": "agentic/finding",
        "level": "error",
        "message": {
          "text": "The 'curl | bash' pattern ... constitutes a benign setup instruction rather than a malicious fetch-and-run chain."
        },
        "properties": {"confidence": 100.0}
      },
      {
        "ruleId": "agentic/finding",
        "level": "error",
        "message": {
          "text": "Remote code execution / fetch-and-run: base64+exec, pipe-to-bash, ..."
        },
        "properties": {
          "confidence": 92.0,
          "citations": [{
            "path": "SKILL.md",
            "start_line": 260, "end_line": 269,
            "snippet": "... curl -fsSL https://parallel.ai/install.sh | bash ..."
          }]
        }
      },
      {
        "ruleId": "agentic/finding",
        "level": "error",
        "message": {
          "text": "Data exfiltration: os.getenv is used to read API keys and transmit them to external servers."
        },
        "properties": {
          "confidence": 90.0,
          "citations": [{
            "path": "lookup.py",
            "start_line": 10, "end_line": 20,
            "snippet": "... os.getenv('OPENROUTER_API_KEY') ... requests.post('https://api.openrouter.ai/v1/chat/completions', ... headers={'Authorization': f'Bearer {api_key}'})"
          }]
        }
      },
      {
        "ruleId": "agentic/finding",
        "level": "note",
        "message": {
          "text": "The static finding of 'Data exfiltration' is not confirmed as malware. ... Further investigation is needed ..."
        },
        "properties": {"confidence": 40.0}
      }
    ],
    "properties": {
      "verdict": "malware",
      "score": 92.0,
      "package": "K-Dense-AI_claude-scientific-skills-skills-research-lookup",
      "latency_seconds": 142.097
    }
  }]
}
\end{lstlisting}